# Martensitic Transformation Precursors: Phonon Theory and Critical Experiments

Yu U. Wang† and Yongmei M. Jin‡
Department of Materials Science and Engineering
Michigan Technological University, Houghton, MI 49931, USA

**Abstract**

This paper addresses martensitic precursor phenomena from new theoretical and experimental perspectives. Driven by a phonon domain hypothesis, based on the premise of incomplete phonon softening, and employing Born's dynamical approach to statistical mechanics of anharmonic crystal lattices, we develop a Grüneisen-type phonon theory that predicts a pre-martensitic transition via "spinodal decomposition" of phonon populations and, without resort to static defects, explains the precursor "anomalies" on the same physical footing of thermal expansion, both being intrinsic properties of anharmonic crystal lattices. The theory reveals the nature of this pre-martensitic transition as phonon pseudo-Jahn-Teller instability transition, and predicts formation of elastic and phonon domains whose behaviors are primarily characterized by the broken dynamic symmetry of lattice vibrations rather than the broken static symmetry of crystal lattice and, as a particular case of Le Chatelier's principle, an internal relaxation in response to applied stress through phonon population redistribution in the domains. The predictions are critically tested by in-situ three-dimensional phonon diffuse scattering and Bragg reflection using high-energy synchrotron X-ray single-crystal diffraction, which observes exotic domain phenomenon in pre-martensitic austenite that is fundamentally different from the usual ferroelastic domain switching phenomenon in martensitic phase. The theory consistently explains the martensitic precursor "anomalies" from the viewpoint of phonon domains.

**Keywords:** Martensitic precursor, Incomplete phonon softening, Phonon domain, Grüneisen-type theory, Phonon spinodal decomposition, Phonon pseudo-Jahn-Teller instability, Pre-martensitic transition, Broken dynamic symmetry of lattice vibrations

## 1. Introduction

Martensitic transformation is a typical solid-state displacive phase transition that breaks crystal symmetry by development of spontaneous anisotropic lattice strain upon cooling [1-3]. Yet, our fundamental understanding of this important phase transformation is still incomplete. Martensitic precursor phenomena remain critical outstanding issues that cannot be well explained from the existing phase transition theories [4]. High-symmetry parent austenite phases usually undergo incomplete phonon softening in a wide temperature range 10-100 K above martensite start temperatures, which is accompanied by various anomalies that are unexpected in cubic austenite phases, including diffuse scattering (streaks and satellites) in diffraction, cross-hatched nanoscale striation image contrast (tweed patterns) in transmission electron microscopy (TEM), and anomalous thermal, acoustic, elastic properties (anisotropic thermal expansion, increased acoustic attenuation, frequency-dependent elastic moduli, etc.) [5-7]. At the heart of the precursor problem is that these anomalies cannot be critical fluctuations because of the exceedingly long lifetime, first-order nature of the martensitic transformation, and high temperature beyond vicinity of the transformation [3,5]. The current explanations of the precursor anomalies (in particular, tweed pattern) are based on static defects (compositional disorder and point defects) that induce strain glass state analogous to spin glass [5,8]. However, in order to consistently explain these anomalies, in particular, the diffuse scattering satellites and their reversible dependences on temperature and stress, implausible ad hoc assumptions have to be made about ordered defect distribution and its reversible evolution and rapid response to stress (as discussed later), making the current explanations questionable.

---

† wangyu@mtu.edu
‡ ymjin@mtu.edu

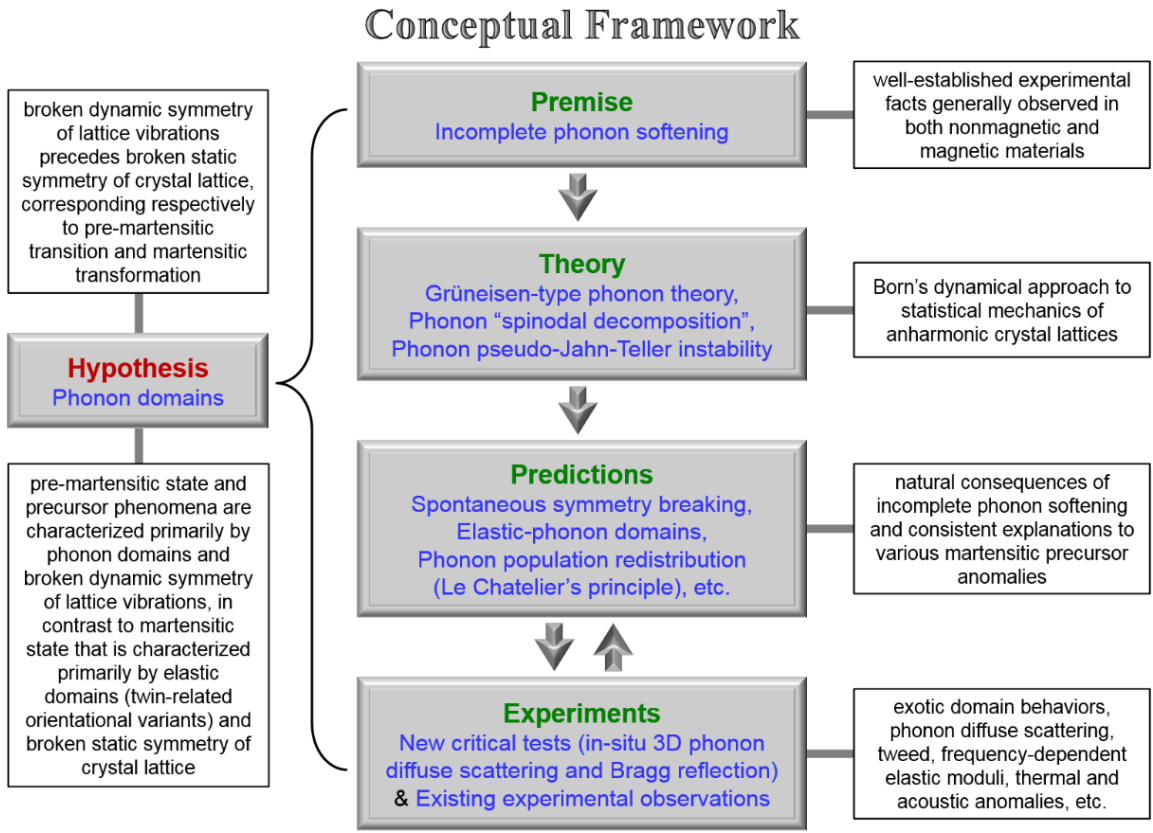

**Figure 1.** Summary of phonon domain hypothesis and the conceptual framework and logical reasoning of the phonon theory of martensitic transformation precursors and related experiments.

In this work, we develop a "simple" theory [9] that provides natural and consistent explanations to martensitic precursor phenomena, and carry out critical experiments to test the theoretical predictions. With various parts of this work reported elsewhere [10-13], this paper presents an overview of this still ongoing work. Figure 1 illustrates the conceptual framework of hypothesis, logical reasoning (premise, theory, prediction), and experimental tests. Driven by a phonon domain hypothesis, based on the premise of incomplete phonon softening, and employing Born's dynamical approach to statistical mechanics of anharmonic crystal lattices, we develop a Grüneisen-type phonon theory [13] that predicts a pre-martensitic transition via "spinodal decomposition" of phonon populations and, without resort to static defects, explains the precursor "anomalies" on the same physical footing of thermal expansion, both being intrinsic properties of anharmonic crystal lattices. The theory further reveals the nature of this pre-martensitic transition as phonon pseudo-Jahn-Teller instability transition [14]. The theory predicts formation of elastic and phonon domains whose behaviors are primarily characterized by the broken dynamic symmetry of lattice vibrations rather than the broken static symmetry of crystal lattice [12] and, as a particular case of Le Chatelier's principle, an internal relaxation in response to applied stress through phonon population redistribution in the domains. The predictions are critically tested by in-situ three-dimensional phonon diffuse scattering and Bragg reflection using high-energy synchrotron X-ray single-crystal diffraction [10-12], which observes exotic domain phenomenon in pre-martensitic austenite that is fundamentally different from the usual ferroelastic domain switching phenomenon in martensitic phase. The theory consistently explains the martensitic precursor "anomalies" from the viewpoint of phonon domains, including TEM tweed patterns, diffraction satellite peaks, and other anomalous properties. The following sections address the key ideas of this work.

## 2. Phonon Theory

### 2.1. Statistical mechanics of anharmonic phonon system

Following Born's dynamical approach to crystal lattices [15], we consider the effective potential function $\Phi$ of a cubic crystal. Given the drastically distinct time scales of lattice deformation and thermal vibrations, the internal coordinates of the crystal can be separated into lattice strain $\varepsilon_{ij}$ (quasi-static configurational coordinates) and atomic displacements $u_\alpha^\mathbf{r}$ (dynamic vibrational coordinates) around equilibrium positions $\mathbf{r}$ of the lattice defined by $\varepsilon_{ij}$. Taking into account the equilibrium condition of undeformed lattice and the invariance relations associated with lattice periodicity and rigid-body translation, the potential $\Phi(\varepsilon_{ij}, u_\alpha^\mathbf{r})$ can be expanded into Taylor series in terms of $\varepsilon_{ij}$ and $u_\alpha^\mathbf{r}$ [16]:

$$\Phi = \Phi_0 + \frac{1}{2} V C_{ijkl}^0 \varepsilon_{ij}\varepsilon_{kl} + \frac{1}{2}\sum_{\mathbf{r},\mathbf{r}'} \mathrm{K}_{\alpha\beta}^0(\mathbf{r}-\mathbf{r}')u_\alpha^\mathbf{r} u_\beta^{\mathbf{r}'} + \frac{1}{2}\sum_{\mathbf{r},\mathbf{r}'} \Delta\mathrm{K}_{\alpha\beta}(\mathbf{r}-\mathbf{r}',\varepsilon_{ij})u_\alpha^\mathbf{r} u_\beta^{\mathbf{r}'} \tag{1}$$

where $\Phi_0 = \Phi(\varepsilon_{ij}=0, u_\alpha^\mathbf{r}=0)$ is a constant term that can be omitted, $C_{ijkl}^0 = V^{-1}\partial^2\Phi/\partial\varepsilon_{ij}\partial\varepsilon_{kl}$ is adiabatic elastic modulus tensor, $V$ is crystal volume, $\mathrm{K}_{\alpha\beta}^0(\mathbf{r}-\mathbf{r}') = \partial^2\Phi/\partial u_\alpha^\mathbf{r}\partial u_\beta^{\mathbf{r}'}$ is Born-von Kármán force constant matrix of the undeformed crystal, $\Delta\mathrm{K}_{\alpha\beta}(\mathbf{r}-\mathbf{r}',\varepsilon_{ij}) = 2(\partial^3\Phi/\partial\varepsilon_{ij}\partial u_\alpha^\mathbf{r}\partial u_\beta^{\mathbf{r}'})\varepsilon_{ij}$ is the perturbation to $\mathrm{K}_{\alpha\beta}^0(\mathbf{r}-\mathbf{r}')$ due to phonon-strain coupling, and all partial derivatives are evaluated at $\varepsilon_{ij} = 0$ and $u_\alpha^{\mathbf{r}} = 0$. The expansion in Eq. (1) is truncated after quadratic terms except the first nonvanishing coupling term between $\varepsilon_{ij}$ and $u_\alpha^{\mathbf{r}}$.

The first quadratic term in Eq. (1) characterizes the elastic energy density of lattice deformation, i.e., $e_\mathrm{L} = \tfrac{1}{2} C_{ijkl}^0 \varepsilon_{ij} \varepsilon_{kl}$. The second quadratic term characterizes the thermal vibrations, and the coupling term accounts for the interactions between the vibrations and strain (i.e., anharmonicity). The eigen-frequencies $\omega_{\mathbf{k},p}$ of normal mode vibrations (phonons) are determined by the secular equation: $\left| \tilde{D}_{\alpha\beta}(\mathbf{k},\varepsilon_{ij}) - \omega_{\mathbf{k},p}^2 \delta_{\alpha\beta} \right| = 0$, where $\tilde{D}_{\alpha\beta}(\mathbf{k},\varepsilon_{ij})$ is the Fourier coefficients of the dynamical matrix $D_{\alpha\beta}(\boldsymbol{\rho},\varepsilon_{ij}) = m^{-1} \mathrm{K}_{\alpha\beta}(\boldsymbol{\rho},\varepsilon_{ij})$, $\mathrm{K}_{\alpha\beta}(\boldsymbol{\rho},\varepsilon_{ij}) = \mathrm{K}_{\alpha\beta}^0(\boldsymbol{\rho}) + \Delta \mathrm{K}_{\alpha\beta}(\boldsymbol{\rho},\varepsilon_{ij})$, $m$ is atomic mass, and the phonon modes are labeled by wavevector $\mathbf{k}$ and polarization $p$. Due to anharmonicity, the phonon frequencies depend on strain, i.e., $\omega_{\mathbf{k},p}(\varepsilon_{ij})$.

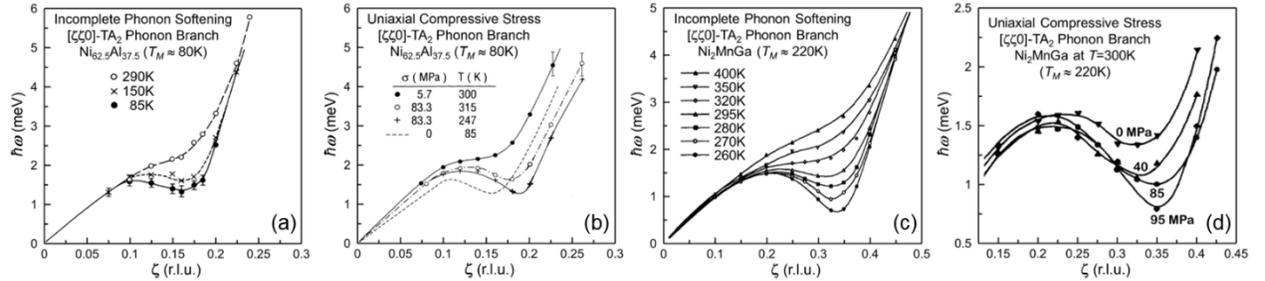

**Figure 2.** Incomplete phonon softening of [ζζ0]-TA$_2$ branch and dependence on (a,c) temperature and (b,d) uniaxial compressive stress in (a,b) nonmagnetic Ni$_{62.5}$Al$_{37.5}$ and (c,d) magnetic Ni$_2$MnGa prior to martensitic transformation [17-20].

To be specific, we consider [ζζ0] transverse acoustic TA$_2$ phonons which are relevant to most martensitic systems [7]. This phonon branch undergoes incomplete softening in austenite phases and exhibits strong deformation dependence, as exemplified in Fig. 2 [17-20]. Such general experimental facts serve as the premise of our theory (it is worth noting that, while the origin of incomplete phonon softening is beyond the scope of our work, the fact that the same phonon phenomenon occurs in both nonmagnetic and magnetic materials indicates that magnetoelastic coupling [21] is not a general origin). It is the deformation-dependent low-energy phonon modes corresponding to the dip in dispersion curve at specific short wavelength that play a dominant role in the lattice stability of cubic austenitic crystals. There are 12 equivalent [ζζ0]-TA$_2$ phonon modes at the dispersion dip in cubic crystal, whose frequency exhibits different dependences on symmetry-breaking anisotropic strain. As illuminating examples, we consider the dependences of phonon frequencies on rhombohedral (trigonal) and tetragonal strains [13,14] (while the theory is general and not limited to the exemplary cases). The rhombohedral (trigonal) distortion is characterized by two independent strain components, namely, volume strain $\varepsilon_V = \varepsilon_1 + \varepsilon_2 + \varepsilon_3$ (where $\varepsilon_1 = \varepsilon_2 = \varepsilon_3$) and shear strain $\gamma = 2\varepsilon_{23} = 2\varepsilon_{31} = 2\varepsilon_{12}$. According to $\mathbf{k}$ vector perpendicular to [111] or not, the 12 [ζζ0]-TA$_2$ phonon modes fall into 2 groups, each group consisting of 6 phonon modes with same frequency dependence on the strain:

$$\omega_1 = \omega_0 + a\varepsilon_V + b\gamma, \quad \omega_2 = \omega_0 + a\varepsilon_V - b\gamma \tag{2}$$

where $\omega_0$ is the frequency in undeformed cubic crystal, and $a$, $b$, $c$ are material constants that characterize the phonon-strain coupling in cubic crystal. The tetragonal distortion is also characterized by two independent strain components, namely, volume strain $\varepsilon_V = \varepsilon_1 + \varepsilon_2 + \varepsilon_3$ (where $\varepsilon_1 = \varepsilon_2 \neq \varepsilon_3$) and tetragonal strain $\eta = \varepsilon_1 - \varepsilon_3 = \varepsilon_2 - \varepsilon_3$. According to **k** vector perpendicular to [001] or not, the 12 [ζζ0]-TA$_2$ phonon modes fall into 2 groups, each consisting of 8 and 4 phonon modes, respectively:

$$\omega_1 = \omega_0 + a\varepsilon_V + c\eta, \quad \omega_2 = \omega_0 + a\varepsilon_V - 2c\eta \tag{3}$$

## 2.2. Grüneisen-type phonon theory of pre-martensitic transition [13]

Above section formulates the statistical mechanics of anharmonic phonon system, based on which the phonon theory of martensitic precursors is developed in this section. For a crystal characterized by the effective potential function $\Phi$ in Eq. (1), the total free energy at temperature $T$ is a sum of lattice free energy $F_L = E_L - TS_L$ and phonon free energy $F_P = E_P - TS_P$. The entropy $S_L$ describes the configurational entropy of a crystal lattice, whose change is negligibly small during diffusionless processes (such as elastic deformation and displacive transformation), thus can be regarded as a constant and omitted in isothermal analysis. The lattice free energy is essentially the elastic energy $E_L = e_L V$. The phonon energy $E_P$ and entropy $S_P$ can be approximated from Planck distribution of independent harmonic oscillators, which give the phonon free energy from all modes $\{\mathbf{k}, p\}$. Normalized by the crystal volume, the total free energy density becomes:

$$F = \frac{1}{2}C_{ijkl}^0 \varepsilon_{ij}\varepsilon_{kl} + \frac{k_B T}{V}\sum_{\mathbf{k}}{}'\sum_{\{\mathbf{k}\}}\sum_p f(x_{\mathbf{k},p}) \tag{4}$$

where $\sum_{\mathbf{k}}'$ indicates sum over the wavevectors **k** only in irreducible first Brillouin zone (IFBZ), $\sum_{\{\mathbf{k}\}}$ indicates sum over all wavevectors $\{\mathbf{k}\}$ that are symmetry-related to the wavevector **k** in IFBZ, $x_{\mathbf{k},p} = \hbar\omega_{\mathbf{k},p}/k_B T$ is dimensionless phonon energy, $f(x) = x/2 + \ln(1-e^{-x})$, $\hbar$ and $k_B$ are reduced Planck constant and Boltzmann constant, respectively. Since it is the deformation-dependent low-energy phonon modes that play the dominant role in the lattice stability of cubic austenitic crystals, $\sum_{\mathbf{k}}' \sum_p$ sum only over a small part of IFBZ near the dispersion dip of [ζζ0]-TA$_2$ phonon modes. For simplicity without loss in conceptual generality, we adopt Einstein-type model of density of states to focus on $N'$ [ζζ0]-TA$_2$ phonon modes around the dispersion dip in first Brillouin zone (FBZ) sharing common frequency $\omega_0$ (or energy $x_0 = \hbar\omega_0/k_B T$) and strain dependence ($N' \ll 3N$, $N$ being the total number of wavevectors in FBZ). This simplification allows an illustration of the basic features determined only by symmetry and a few material parameters ($a$, $b$, $c$).

The volume strain $\varepsilon_V$ characterizes symmetry-preserving thermal expansion phenomenon (as in Grüneisen model) and is determined by $(\partial F/\partial \varepsilon_V)_{\eta=0} = 0$, which yields $\varepsilon_V^0 = -9\chi\hbar f'(x_0)a/\Omega(C_{11}+2C_{12})$, where $\Omega = V/N$ is primitive cell volume, and $\chi = N'/3N$ (note that $\varepsilon_V^0 = \varepsilon_V^0[\omega_0(T), T]$ is temperature dependent). The shear strain $\gamma$ and tetragonal strain $\eta$ characterize symmetry-breaking anisotropic lattice distortions. Because $\partial F/\partial \gamma \equiv 0$ and $\partial F/\partial \eta \equiv 0$ for cubic lattice, the cubic crystal loses lattice stability against rhombohedral (trigonal) and tetragonal distortions under the following respective conditions:

$$\frac{\partial^2 F}{\partial \gamma^2} = 3\left[C_{44}^0 + \chi\frac{\hbar^2}{\Omega k_B T}f''(x_0')b^2\right] < 0, \quad \frac{\partial^2 F}{\partial \eta^2} = \frac{2}{3}\left[(C_{11}^0 - C_{12}^0) + 9\chi\frac{\hbar^2}{\Omega k_B T}f''(x_0')c^2\right] < 0 \quad (5)$$

where $x_0'$ (or $\omega_0'$) incorporates the effect of thermal expansion. Eq. (5) describes spinodal instability since it is related to the curvature (second derivative) of the free energy in analogy to spinodal decomposition. Figure 3(a) plots $f(x)$, $f'(x)$ and $f''(x)$. With decreasing $x = \hbar\omega/k_B T$ (i.e., increasingly softened phonons with decreasing temperature), $f''(x)$ rapidly approaches $-\infty$ and, as a result, the instability conditions in Eq. (5) are achieved without requirement of soft mode (i.e., phonon frequency does not vanish).

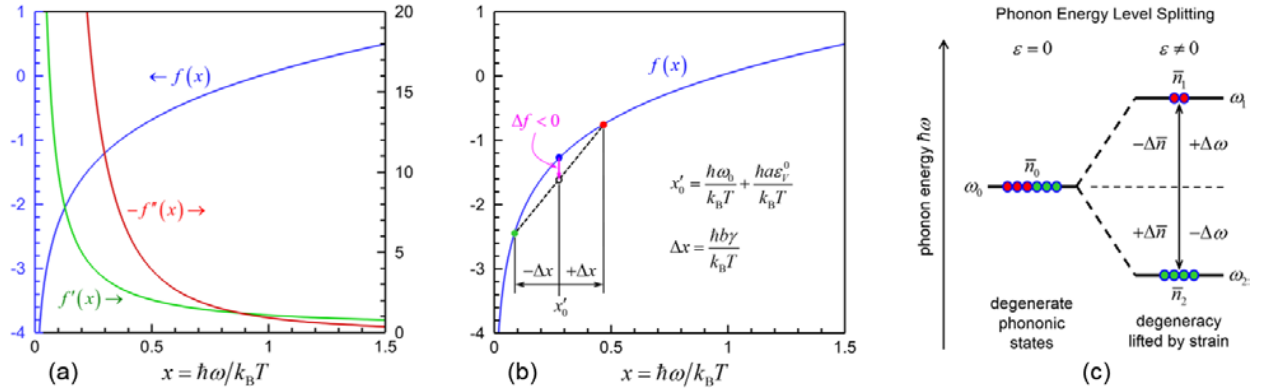

**Figure 3.** (a) Dependences of $f(x)$, $f'(x)$ and $f''(x)$ on $x = \hbar\omega/k_B T$. (b) Illustration of pre-martensitic transition via "spinodal decomposition" of phonon populations associated with lattice distortion strain. (c) Schematic of phonon energy level splitting ($\omega$) and phonon population redistribution ($\bar{n}$) due to lifting of phonon degeneracy by lattice strain ($\varepsilon$).

Above phonon theory predicts a pre-martensitic transition via "spinodal decomposition" of phonon populations, as illustrated in Fig. 3(b) for the case of rhombohedral (trigonal) strain. Symmetry-breaking strain induces phonon energy changes, as exemplified in Eqs. (2) and (3), and phonon population redistributions; because $f''(x) < 0$, the total phonon free energy decreases; thus, the phonon populations have an intrinsic tendency to "decompose" into elastic-phonon domains. This tendency is resisted by elastic energy. Under condition specified in Eq. (5), the phonon free energy decrease outweighs the elastic energy increase, leading to pre-martensitic transition. This transition occurs at finite phonon frequency of ~1 THz (corresponding to ~1 meV energy of incompletely softened phonon). It is worth noting that the phonon dispersion in the cubic austenite and pre-martensitic phase exhibits a continuous dependence on temperature across the pre-martensitic transition, which is in contrast to the discontinuous change in phonon dispersion during martensitic transformation.

## 2.3. Phonon pseudo-Jahn-Teller instability transition [14]

The nature of the pre-martensitic transition via "spinodal decomposition" of phonon populations can be interpreted as a phonon pseudo-Jahn-Teller instability transition, which is caused by the interactions between fast lattice vibrations and slow lattice deformation (i.e., phonon-strain coupling), in analogy to Jahn-Teller instability caused by electron-phonon coupling. During the pre-martensitic transition, the crystal lattice spontaneously deforms to break the cubic symmetry and lift the phonon degeneracy, and in so doing lower the overall phonon energy. The phonon energy is $E_\text{P} = \sum_{\mathbf{k},p}(\tfrac{1}{2}+\bar{n}_{\mathbf{k},p})\hbar\omega_{\mathbf{k},p}$, where $\bar{n}_{\mathbf{k},p} = [\exp(\hbar\omega_{\mathbf{k},p}/k_\text{B}T)-1]^{-1}$ is the thermal equilibrium occupation number of phonons with energy $\hbar\omega_{\mathbf{k},p}$. For cubic lattice, the above considered 12 equivalent [ζζ0]-TA$_2$ phonon modes are degenerate with the same frequency $\omega_0$ and equilibrium occupation number $\bar{n}_0$ for each mode. Such a degenerate phononic state is unstable due to phonon-strain coupling. Under rhombohedral (trigonal) lattice strain, the 12 phonon modes fall into 2 groups (6 in each) that undergo phonon energy level splitting: $\omega_1 = \omega_0 + \Delta\omega$ and $\omega_2 = \omega_0 - \Delta\omega$, as shown in Eq. (2) and illustrated in Fig. 3(c). Consequently, the phonon populations redistribute: $\bar{n}_1 = \bar{n}_0 - \Delta\bar{n}$ and $\bar{n}_2 = \bar{n}_0 + \Delta\bar{n}$, where

$$\Delta\bar{n} = \frac{\Delta\omega}{\omega_0}\frac{\hbar\omega_0}{k_\text{B}T}\exp\left(\frac{\hbar\omega_0}{k_\text{B}T}\right)\left[\exp\left(\frac{\hbar\omega_0}{k_\text{B}T}\right)-1\right]^{-2} \quad (6)$$

Lifting phonon degeneracy lowers overall phonon energy: $\Delta E_\text{P} = -12\hbar\Delta\bar{n}\Delta\omega < 0$. Similarly, under tetragonal lattice strain, the 12 phonon modes fall into 2 groups (8 in one and 4 in the other) that undergo phonon energy level splitting, $\omega_1 = \omega_0 + \Delta\omega$, $\omega_2 = \omega_0 - 2\Delta\omega$, as shown in Eq. (3); consequently, the phonon populations redistribute, $\bar{n}_1 = \bar{n}_0 - \Delta\bar{n}$, $\bar{n}_2 = \bar{n}_0 + 2\Delta\bar{n}$; and lifting phonon degeneracy lowers overall phonon energy, $\Delta E_\text{P} = -24\hbar\Delta\bar{n}\Delta\omega < 0$. It is worth noting that $\Delta E_\text{P} \propto -\Delta\bar{n}\Delta\omega \propto -\varepsilon^2$ is a general feature, because $\Delta\omega \propto \varepsilon$ as shown in Eqs. (2) and (3) and $\Delta\bar{n} \propto \Delta\omega$ as shown in Eq. (6); thus the phonon energy reduction is quadratically proportional to strain and competes with the lattice strain energy. The strain-induced phonon energy level splitting and phonon population redistribution are equivalent to the "spinodal decomposition" of phonon populations.

The phonon-strain pseudo-Jahn-Teller instability exhibits some features distinct from the electron-phonon Jahn-Teller instability. Unlike electrons being fermions thus degenerate orbitals must be partially filled to produce Jahn-Teller instability, phonons are boson quasiparticles that are free to redistribute, thus phonon-strain coupling is a general mechanism of lattice instability. Because $\Delta \bar{n} \propto \Delta \omega \propto \varepsilon$, the energy decrease in phonon pseudo-Jahn-Teller effect is quadratically proportional to strain (i.e., $\Delta E_\mathrm{P} \propto -\Delta \bar{n} \Delta \omega \propto -\varepsilon^2$), while in Jahn-Teller effect the energy decrease is linearly proportional to strain since the filling of lower-energy orbitals in ground state does not depend on the strain. The strength of such phonon-strain coupling as measured by $\Delta E_\mathrm{P}$ depends not only on the material parameters $a$, $b$, $c$ but also, more importantly, on the low phonon energy $\hbar \omega_0$. Incomplete phonon softening provides the required low phonon energy to exhibit the phonon pseudo-Jahn-Teller instability pre-martensitic transition.

## 2.4. Predictions

The central prediction of above phonon theory is formation of phonon domains, confirming the phonon domain hypothesis. The pre-martensitic transition breaks the cubic crystal symmetry and leads to formation of elastic domains. Elastic domains are characterized by the time-averaged static symmetry of crystal lattice, whose strength is characterized by the spontaneous anisotropic strain (e.g., $\gamma$ and $\eta$). The spontaneous strain of pre-martensitic phase is, however, very small as compared with the transformation strain of martensitic phase, thus its measurement requires high-resolution single-crystal diffraction, while pre-martensitic phase remains quasi-cubic in conventional diffraction. Most importantly, the elastic domains are also phonon domains of redistributed phonon populations due to "spinodal decomposition." Phonon domains are characterized by the broken dynamic symmetry of lattice vibrations [12], whose strength is characterized by

$$\frac{\Delta \bar{n}}{\bar{n}_0} = \frac{\Delta \omega}{\omega_0} \frac{\hbar \omega_0}{k_\mathrm{B} T} \exp\left(\frac{\hbar \omega_0}{k_\mathrm{B} T}\right) \left[\exp\left(\frac{\hbar \omega_0}{k_\mathrm{B} T}\right) - 1\right]^{-1} \qquad (7)$$

The properties of pre-martensitic phase are primarily characterized by the broken dynamic symmetry of lattice vibrations (phonon domains) rather than the broken static symmetry of crystal lattice (elastic domains). Each elastic-phonon domain contains a dominant phonon mode of lowered phonon energy due to strain-induced energy level splitting. As a particular case of Le Chatelier's principle, the phonon domain concept predicts an internal relaxation through phonon population redistribution in the domains in response to applied stress. This extra degree of freedom allows pre-martensitic crystal to respond to applied stress via domain deformation, in drastic contrast to usual ferroelastic domain switching in martensitic phase (such exotic domain phenomenon is confirmed by critical experiment, as discussed in Section 3.2). The relaxation time of phonon population redistribution leads to frequency dependence of acoustic wave propagation in pre-martensitic crystal; in particular, it is predicted that much higher elastic constants are measured by high-frequency ultrasound than by low-frequency (with longer period relative to phonon relaxation time) and quasi-static measurements, which provides a means to determine the phonon relaxation time (it is worth noting that such frequency dependence has been observed in experiments: for example in pre-martensitic Ni-Mn-Ga, ultrasonic wave velocity measurement reports elastic modulus ~150 GPa [22], while three-point bending measurement at 1 Hz reports elastic modulus ~20 GPa [23]).

The phonon domain concept explains the martensitic precursor "anomalies." The diffuse streaks and satellites in diffraction originate from phonon diffuse scattering [10-12], as illustrated in Fig. 4(a,b) and experimentally observed in Fig. 4(c-f). In particular, when a negative dip develops in phonon dispersion curve upon cooling, phonon diffuse scattering produces a satellite peak accordingly, which has been generally misinterpreted as evidence of static lattice modulations (modulated phase). Such diffuse satellites are produced by scattering from incompletely softened phonons of typical energy ~1 meV and frequency ~1 THz, thus there is no diffraction evidence that lattice modulations or modulated phase exist in pre-martensitic phase. It is worth noting that phonon domains produce diffraction contrast and tweed patterns in TEM imaging, as illustrated in Fig. 4(g). Phonon domains are localized lattice vibrations of dominant modes [12]. Electron diffraction probes instantaneous atomic positions of vibrating lattice, producing instantaneous diffraction image contrast. While traveling waves carry such instantaneous image contrast to different regions thus averaging out the contrast, localized lattice vibrations within phonon domains consistently produce such diffraction image contrast in the same lattice regions that adds up to produce tweed patterns. The anomalous thermal expansion and acoustic attenuation in the pre-martensitic crystal are expected from the structural heterogeneities caused by the formation of elastic-phonon domains.

## 3. Critical Experiments

Above predicted temperature- and stress-dependent behaviors of elastic-phonon domains are critically tested by in-situ three-dimensional phonon diffuse scattering and Bragg reflection using high-energy synchrotron X-ray single-crystal diffraction [10-12], which provides complementary information of crystal lattice and lattice vibrations. Single crystals of $Ni_{49.90}Mn_{28.75}Ga_{21.35}$ (Adaptamat®) are used as model system, which are measured at temperatures above the martensite start temperature ($T_M$=323 K).

### 3.1. Temperature-dependent phonon diffuse scattering

Figure 4(c,e,f) shows the phonon diffuse scattering intensity distribution as function of temperature [12]. As predicted from incomplete phonon softening and illustrated in Fig. 4(a,b), diffuse streaks become stronger upon cooling and form diffuse satellites when a negative dip develops in phonon dispersion curve, both originating from phonon diffuse scattering. The negative dip deepens upon cooling, which leads to more phonon concentration (the thermal average number of phonons $n(k)$ obeys Bose-Einstein statistics and concentrates at the wavevector of the dip), stronger diffuse satellites (the phonon amplitude $u(k)$ is related to $n(k)$ through $u^2 \propto (n+\frac{1}{2})/\omega$, which accordingly develops a peak, and phonon diffuse scattering intensity is $I \propto [(\mathbf{G}+\mathbf{k})\cdot\mathbf{e}]^2 u^2$, $\mathbf{k}$, $\mathbf{e}$ and $\mathbf{G}$ being phonon wavevector, polarization vector and reciprocal lattice vector corresponding to Bragg reflection), and increased diffraction image contrast of tweed patterns (it is worth noting that "tweed" is a description of image contrast in TEM rather than a definition of lattice structure, thus the tweeds of martensitic precursor and coherent precipitation, e.g., G.P. zones formed by atomic diffusion, must be distinguished). The apparent incommensurability in the satellite position and its evolution with temperature correspond to the wavevector at the negative dip in incompletely softened phonon dispersion. Because $I \propto \mathbf{G}\cdot\mathbf{e}$ (since $\mathbf{k}\cdot\mathbf{e}=0$ for $[\zeta\zeta 0]$-$TA_2$ phonons), phonon diffuse scattering exhibits systematic extinction, as shown in Fig. 4(d).

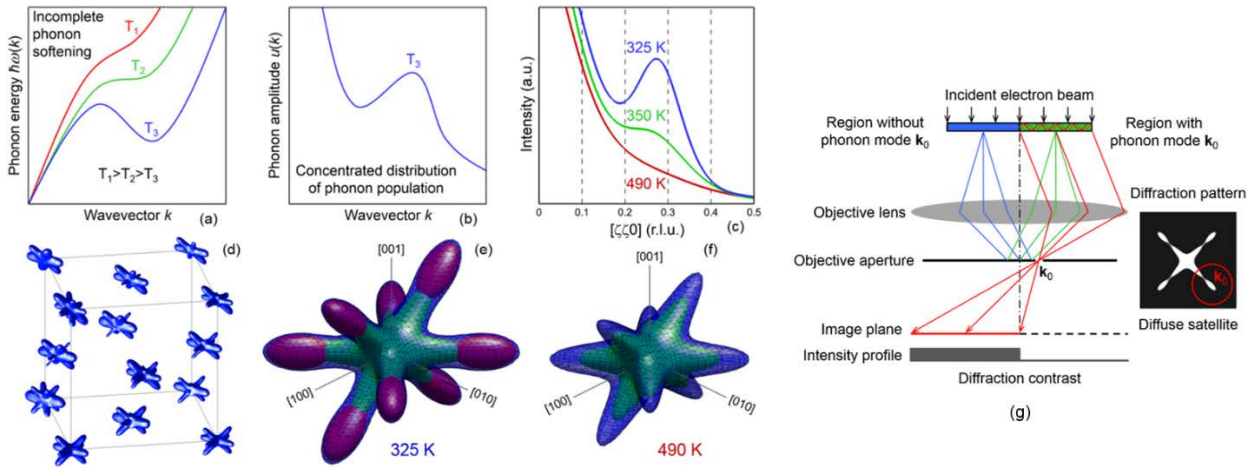

**Figure 4.** (a) Phonon dispersion at temperatures $T_1 > T_2 > T_3 > T_M$; a negative dip develops at $T_3$. (b) Distribution of phonon population (phonon amplitude) concentrates into a peak (corresponding to the dip in dispersion) at $T_3$. (c) Experimental measurement of temperature-dependent phonon diffuse scattering intensity distribution; a diffuse satellite peak develops at 325 K. (d) 3D phonon diffuse scattering around different Bragg reflections without applied stress. (e,f) 3D phonon diffuse scattering around (800) Bragg reflection at 325K with satellites and 490 K without satellites. (g) Diffraction contrast imaging of phonon domains in TEM.

### 3.2. Exotic domain phenomenon under stress

To detect the behaviors of elastic-phonon domains, three-dimensional phonon diffuse scattering and Bragg reflection are measured under in-situ uniaxial stress applied along single crystal [001] axis at 350K ($T_M$=323 K). Figure 5(a,b) shows the domain phenomenon identified from Bragg reflection. The most unusual behavior is that the domains deform into two (soft and hard) tetragonal lattices but do not switch. In particular, the volume fractions of soft and hard domain variants remain ⅓ and ⅔, respectively, during entire loading-unloading cycle, which reflect the volume fractions of the phonon domains as determined by $\mathbf{k}$ vectors perpendicular to [001] or not (i.e., 4:8) as discussed in Eq. (3). The lattice parameters as functions of stress exhibit hysteresis-free reversibility, as shown in Fig. 5(g) where solid-filled symbols represent unloading, indicating elastic deformation behaviors of both domains. The lattice parameters $a_{s[100]}$ and $a_{s[001]}$ respectively along single crystal [100] and [001] axes correspond to the elastically deformed soft domains with **k** along [110] or [1$\bar{1}$0], while $a_{h[100]}$ and $a_{h[001]}$ correspond to the hard domains with **k** along [011], [01$\bar{1}$], [101], or [10$\bar{1}$]. Figure 5(c-e) shows that the in-plane diffuse scattering corresponding to **k** perpendicular to [001]-stress (**k** along [110] and [1$\bar{1}$0] within yellow plane) grows in intensity with increasing stress, while the out-of-plane diffuse scattering is depressed by the stress. The in-plane diffuse satellites not only grow in intensity, but also concentrate into narrower peaks, and their incommensurate positions gradually shift towards $\zeta=⅓$ with increasing stress, as shown in Fig. 5(f). These observations agree with the stress-enhanced phonon softening behaviors directly observed by inelastic neutron scattering in Fig. 2 [17-20]. To confirm that each of the two tetragonal domains produces own phonon diffuse scattering, the lattice parameter $a$ is also determined respectively from the centers of in-plane and out-of-plane diffuse scatterings around (800) Bragg peak, as plotted in black symbols and lines in Fig. 5(g). The out-of-plane diffuse scattering gives the same $a$ as $a_{h[100]}$, while the in-plane diffuse scattering gives $a$ that is close to $a_{h[100]}$ at low stress and approaches $a_{s[100]}$ with increasing stress. This analysis result is caused by the fact that the in-plane diffuse scattering is increasingly dominated by the soft phonon domains with increasing stress, and the out-of-plane diffuse scattering is dominated by the hard phonon domains. It is worth noting that the soft phonon domains exhibit phonon-mediated large, reversible, anhysteretic super-elastic strain response. While the hard phonon domains exhibit Young's modulus $E$=14 GPa and normal Poisson's ratio $\nu$=0.33, the soft domains exhibit significantly lower Young's modulus $E$=4.8 GPa and unusually high Poisson's ratio $\nu$=0.47 (approaching thermodynamic upper limit ½). Such exotic domain behaviors observed by combined in-situ three-dimensional phonon diffuse scattering and Bragg reflection agree with the phonon domain concept and phonon population relaxation (Le Chatelier's principle).

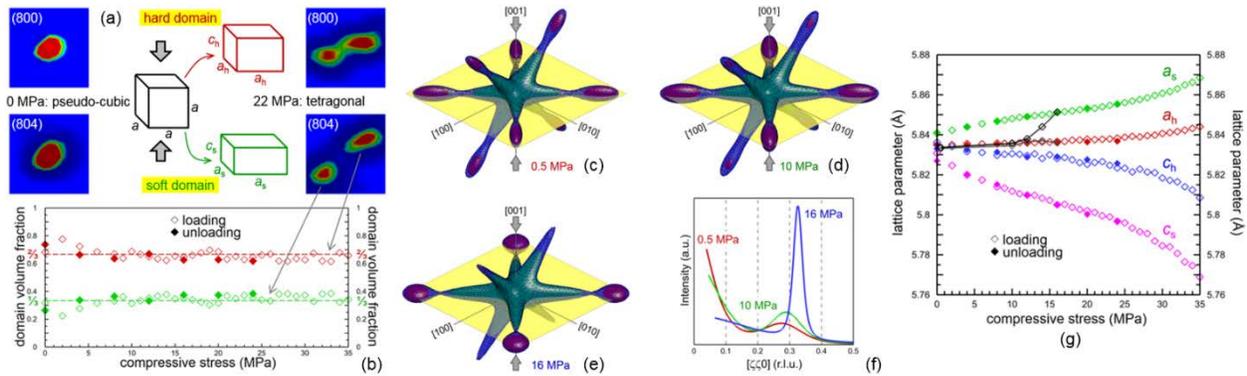

**Figure 5.** (a) Bragg reflection peaks under [001] stress. (b) Conservation of domain variant volume fractions under stress. (c-e) Stress-dependent 3D phonon diffuse scattering around (800) Bragg peak at 350 K ($>T_M$). (f) Stress dependence of diffuse scattering intensity distribution. (g) Stress-dependent lattice parameters of hard (red, blue) and soft (green, pink) phonon domains as determined from Bragg peaks in (a). Lattice parameters (black) are also determined from centers of 3D diffuse scattering in (c-e).

It is worth noting that above observed temperature- and stress-dependent evolutions of diffuse satellites cannot be explained by static defects (compositional disorder, point defects). To explain the reversible dependences of diffuse satellites on temperature and stress, reversible evolution of ordered defect distributions (required to produce the satellites) and their rapid response to stress must be assumed, which are implausible. On the contrary, incomplete phonon softening naturally explains these experimental observations.

## 4. Summary

Driven by a hypothesis of phonon domains, we develop a statistical mechanics theory (rather than phenomenological model) of pre-martensitic transition, which confirms the phonon domain hypothesis, explains the martensitic precursor phenomena, and predicts new domain phenomenon that is successfully tested by critical experiments. The following new picture of martensitic transformation emerges from this theoretical and experimental work: due to incomplete phonon softening in austenite phase, pre-martensitic transition naturally precedes martensitic transformation, where the former leads to elastic-phonon domains that are primarily characterized by broken dynamic symmetry of lattice vibrations while the latter leads to elastic domains that are primarily characterized by broken static symmetry of crystal lattice. As our ongoing research, we are still working to address the origin of incomplete phonon softening, which is the premise and beyond the scope of our theory, but is important to understand its fundamental mechanism.

**Acknowledgements:** We sincerely acknowledge Professor Armen G. Khachaturyan for valuable discussion on Jahn-Teller effect. We are very grateful to Professor Long-Qing Chen for his passionate encouragement and constant support to us during this adventurous effort over the past five years. This work was initially supported by NSF DMR-0800048 and subsequently partially by DOE DE-FG02-09ER46674. Use of Advanced Photon Source was supported by DOE DE-AC02-06CH11357.

**References and Notes**
[1]    Z. Nishiyama, *Martensitic Transformation* (Academic Press, New York, 1978).
[2]    A.G. Khachaturyan, *Theory of Structural Transformations in Solids* (John Wiley & Sons, New York, 1983).
[3]    E.K.H. Salje, *Phase Transitions in Ferroelastic and Co-elastic Crystals* (Cambridge University Press, Cambridge, 1990).
[4]    Just to illustrate how puzzling the precursor phenomena are, former American Physical Society President, late Professor J.A. Krumhansl, regarded "Precursors: The Next Frontier" (title of the 3$^{rd}$ section) and called for renormalization group theory to tackle the precursor problem in his conceptual overview [5] (his last publication searchable in Web of Science); Otsuka and Ren started the 5$^{th}$ section "Precursor Effects to Martensitic Transformations" with an analogy between precursor phenomena prior to martensitic transformation and social unrest ("battles" and "conflicts") before social revolution in their review [6].

[5]  J.A. Krumhansl, "Multiscale Science: Materials in the 21st Century," *Mater. Sci. Forum*, **327-328**, 1-8 (2000).

[6]  K. Otsuka, X. Ren, "Physical Metallurgy of Ti-Ni-based Shape Memory Alloys," *Prog. Mater. Sci.*, **50**, 511-678 (2005).

[7]  N. Nakanishi, "Elastic Constants as They Relate to Lattice Properties and Martensite Formation," *Prog. Mater. Sci.*, **24**, 143-265 (1980).

[8]  S. Kartha, J.A. Krumhansl, J.P. Sethna, L.K. Wickham, "Disorder-Driven Pretransitional Tweed Pattern in Martensitic Transformations," *Phys. Rev. B*, **52**, 803-822 (1995).

[9]  We have been striving to develop a "simple" theory in the sense that John A. Wheeler used to say: "Surely someday, we can believe, we will grasp the central idea of it all as so simple, so beautiful, so compelling that we will say to each other, 'Oh, how could it have been otherwise! How could we all have been so blind so long!' " (J.A. Wheeler, "Information, Physics, Quantum: The Search for Links," in *Complexity, Entropy and the Physics of Information*, edited by W.H. Zurek, Westview Press, Boulder, Colorado, 1990).

[10] T.L. Cheng, F.D. Ma, J.E. Zhou, G. Jennings, Y. Ren, Y.M. Jin, Y.U. Wang, "In-Situ Three-Dimensional Reciprocal-Space Mapping of Diffuse Scattering Intensity Distribution and Data Analysis for Precursor Phenomenon in Shape-Memory Alloy," *JOM*, **64**, 167-173 (2012).

[11] Y.M. Jin, Y. Ren, Y.U. Wang (invited speaker), "Three-Dimensional Diffuse Scattering Study of Phase Transition Precursor Phenomenon Using In-Situ High-Energy X-Ray Single-Crystal Diffraction," May 8, 2013 APS/CNM/EMC Users Meeting, Argonne National Laboratory.

[12] Y.M. Jin, Y.U. Wang, Y. Ren, et al, "Broken Dynamic Symmetry and Phase Transition Precursor," arXiv:1302.5479 (2013).

[13] Y.M. Jin, Y.U. Wang, "Phonon Theory of Martensitic Transformation Precursors," arXiv:1412.3725 (2014).

[14] Y.M. Jin, Y.U. Wang, A.G. Khachaturyan, "Phonon Pseudo-Jahn-Teller Instability Transition" (unpublished).

[15] M. Born, K. Huang, *Dynamical Theory of Crystal Lattices* (Oxford University Press, Oxford, 1954).

[16] For simplicity without loss in conceptual generality, crystal with monatomic basis is formulated here; crystal with polyatomic basis can be treated in similar manner [15].

[17] S.M. Shapiro, B.X. Yang, G. Shirane, Y. Noda, L.E. Tanner, "Neutron Scattering Study of the Martensitic Transformation in a Ni-Al β-Phase Alloy," *Phys. Rev. Lett.*, **62**, 1298-1301 (1989).

[18] S.M. Shapiro, E.C. Svensson, C. Vettier, B. Hennion, "Uniaxial-Stress Dependence of the Phonon Behavior in the Premartensitic Phase of $Ni_{62.5}Al_{37.5}$," *Phys. Rev. B*, **48**, 13223-13229 (1993).

[19] A. Zheludev, S.M. Shapiro, P. Wochner, A. Schwartz, M. Wall, L.E. Tanner, "Phonon Anomaly, Central Peak, and Microstructures in Ni$_2$MnGa," *Phys. Rev. B*, **51**, 11310-11314 (1995).

[20] A. Zheludev, S.M. Shapiro, "Uniaxial Stress Dependence of the [ζζ0]-TA$_2$ Anomalous Phonon Branch in Ni$_2$MnGa," *Solid State Communications*, **98**, 35-39 (1996).

[21] M.A. Uijttewaal, T. Hickel, J. Neugebauer, M.E. Gruner, P. Entel, "Understanding the Phase Transitions of the Ni$_2$MnGa Magnetic Shape Memory System from First Principles," *Phys. Rev. Lett.*, **102**, 035702-1-4 (2009).

[22] M. Stipcich, L. Mañosa, A. Planes, M. Morin, J. Zarestky, T. Lograsso, C. Stassis, "Elastic Constants of Ni-Mn-Ga Magnetic Shape Memory Alloys," *Phys. Rev. B.*, **70**, 054115-1-5 (2004).

[23] V.A. Chernenko, J. Pons, C. Seguí, E. Cesari, "Premartensitic Phenomena and Other Phase Transformations in Ni-Mn-Ga Alloys Studied by Dynamical Mechanical Analysis and Electron Diffraction," *Acta Mater.*, **50**, 53-60 (2002).